\documentclass[%
 reprint,
 amsmath,amssymb,
 aps,
]{revtex4-1}

\usepackage{graphicx}
\usepackage{dcolumn}
\usepackage{natbib}
\usepackage{bm}

\usepackage[toc]{appendix}
\usepackage{graphicx}
\usepackage{epstopdf} 
\usepackage{relsize} 
\usepackage[mathscr]{euscript} 

\begin{document}

\preprint{APS}

\title{Noise and dynamics in forward Brillouin interactions}

\author{P. Kharel}
\email{prashanta.kharel@yale.edu}
\author{R. O. Behunin}
\author{W. H. Renninger}
\author{P. T. Rakich}

\affiliation{%
 Department of Applied Physics, Yale University, New Haven, CT 06511
}%

\date{\today}

\begin{abstract}
In this paper, we explore the spatio-temporal dynamics of spontaneous and stimulated forward Brillouin scattering. This general treatment incorporates the optomechanical coupling produced by boundary-induced radiation pressures (boundary motion) and material-induced electrostrictive forces (photo-elastic effects), permitting straightforward application to a range of emerging micro- and nano-scale optomechanical systems. Through a self-consistent fully coupled nonlinear treatment, developed within a general Hamiltonian framework, we establish the connection between the power spectral density of spontaneously scattered light in forward Brillouin interactions and the nonlinear coupling strength. We show that, in sharp contrast to backward Brillouin scattering, noise-initiated stimulated forward Brillouin scattering is forbidden in the majority of experimental systems. In fact, the single-pass gain, which characterizes the threshold for energy transfer in back-scattering processes, is negative for a large class of forward Brillouin devices.  Beyond this frequent experimental case, we explore mechanisms for dispersive symmetry breaking that lead to amplification and dynamics reminiscent of backward Brillouin scattering.
\end{abstract}

\pacs{42.65.-k, 42.50.-p}
\maketitle


\section{\label{sec:level1} Introduction}

Micro- and nano-scale structural control has been used to enhance and tailor interactions between photons and phonons in a range of new systems \cite{dainese2006stimulated, shibata2006forward, beugnot2007guided, zhu2007stored, kippenberg2008cavity, kang2009tightly, lin2009mechanical, grudinin2010phonon, pant2011chip, butsch2012optomechanical, bahl2012observation, aspelmeyer2014cavity, fiore2013optomechanical, bochmann2013nanomechanical, shin2013tailorable, qiu2013stimulated, shin2015control, van2015interaction, kittlaus2015large}, giving rise to a great diversity of optomechanical interactions \cite{kang2009tightly,lin2009mechanical, vahala2009phonon, grudinin2010phonon, pant2011chip,chan2011laser, safavi2011electromagnetically,butsch2012optomechanical, bahl2012observation,hill2012coherent, aspelmeyer2014cavity, fiore2013optomechanical, bochmann2013nanomechanical,butsch2014cw, shin2015control, kittlaus2015large, pang2015stable}.  These new optomechanical systems provide a powerful interface between optical and phononic domains as the basis for both classical \cite{zhu2007stored, byrnes2012photonic, li2013microwave, marpaung2013frequency, pant2014chip, marpaung2014nonlinear, shin2015control} and quantum \cite{brooks2012non, ludwig2012enhanced, bochmann2013nanomechanical, aspelmeyer2014cavity} signal processing operations.  Among these optomechanical systems are a new class of hybrid photonic-phononic waveguides that permit new engineerable forms of traveling-wave photon-phonon coupling \cite{dainese2006stimulated, shibata2006forward, beugnot2007guided, kang2009tightly, grudinin2010phonon, pant2011chip, butsch2012optomechanical, qiu2013stimulated, shin2015control, van2015interaction, kittlaus2015large}.  These traveling-wave interactions, broadly termed Brillouin interactions, are the basis for tailorable forms of signal amplification \cite{beugnot2007guided, kang2009tightly, grudinin2010phonon, pant2011chip, shin2013tailorable, qiu2012stimulated, qiu2013stimulated, van2015interaction, kittlaus2015large}, high performance lasers \cite{byrnes2012photonic, li2013microwave, li2014low}, and a host of hybrid photonic-phononic signal processing operations that have no optical analog \cite{zhu2007stored, marpaung2013frequency, shin2015control}. Such highly engineerable couplings have given rise to new types and regimes of Brillouin interactions \cite{kang2009tightly, kang2010all, butsch2014cw}, and more complex optomechanical processes that challenge the definition of Brillouin processes \cite{butsch2014cw, conti2012dynamics, shin2015control, van2015unifying}. While Brillouin physics has a rich history \cite{boyd2003nonlinear, agrawal2007nonlinear, damzen2003stimulated}, with the emergence of these enhanced forms of photon-phonon coupling, established models of Brillouin noise and dynamics no longer apply.

These new optomechanical (or Brillouin-active) waveguides, achieve radical enhancement of forward-Brillouin coupling (scattering) through confinement of guided optical and acoustic modes within microstructured fibers and nanophotonic waveguides, providing access to rich new regimes of nonlinear dynamics \cite{dainese2006stimulated, shibata2006forward, beugnot2007guided, kippenberg2008cavity, kang2009tightly, lin2009mechanical, grudinin2010phonon, pant2011chip, shin2013tailorable, qiu2013stimulated, shin2015control, van2015interaction, kittlaus2015large}.  Forward-Brillouin interactions (not to be confused with more widely studied backward Brillouin processes) are characterized by phonon-mediated coupling between co-directionally propagating optical waves \cite{kang2009tightly}, whereas backward-Brillouin interactions couple contra-directionally propagating optical waves \cite{boyd2003nonlinear, agrawal2007nonlinear}. In contrast to backwards Brillouin scattering, wave-guidance is required to achieve phase-matched forward-Brillouin scattering. Moreover, the frequency, strength, and type of coupling is far more tailorable owing to the inherent geometric dependence of forward Brillouin interactions \cite{kang2009tightly, rakich2012giant, shin2013tailorable}. While this new device physics holds much technological promise, little is known about the noise and noise initiated threshold conditions for such interactions.

The noise and dynamics of backward-Brillouin scattering has been extensively studied in the context of fiber optic technologies \cite{boyd1990noise, gaeta1991stochastic, yeniay2002spontaneous, le2003theoretical}.  However, until recently, forward-Brillouin couplings have been very weak by comparison, making their technological importance less apparent. The first systematic studies of forward Brillouin interactions focused on \emph{spontaneous} forward-Brillouin scattering \cite{shelby1985guided}, not to be confused with \emph{stimulated} forward Brillouin scattering \cite{kang2009tightly}. Through these studies Shelby \emph{et al.}, identified spontaneous  forward-Brillouin scattering, also termed guided acoustic wave Brillouin scattering (GAWBS), as a key source of noise in fiber-based quantum optics measurements \cite{shelby1986broad, elser2006reduction}. A theoretical framework was also developed to describe how the phase and polarization noise that thermally driven guided acoustic modes impart to light through photo-elastic coupling in optical fibers \cite{shelby1986broad, elser2006reduction}. However, to capture the noise characteristics of a diversity of new fiber and waveguide geometries \cite{butsch2014cw, conti2012dynamics, rakich2012giant, shin2013tailorable, van2014analysis, van2015interaction}, it is necessary to incorporate both photo-elastic response and boundary motion in a more general formulation of Brillouin noise.  Beyond spontaneous Brillouin noise, little is known about the noise and threshold properties of these interactions, which are important to the development of Brillouin based signal processing technologies.

To address these challenges, we build on the quantum traveling-wave treatment of Brillouin coupling \cite{sipe2015hamiltonian}, and prior quantum-traveling wave treatments of noise and nonlinearity \cite{haus1970steady, imoto1985quantum, blow1990continuum, jeffers1993quantum, haus2000electromagnetic, loudon2000quantum}. This approach captures the distributed optomechanical coupling, noise, and spatio-temporal field evolution in Brillouin interactions within a generalizable Hamiltonian framework. Moreoever, this formulation incorporates the optomechanical couplings produced by boundary-induced radiation pressures (boundary motion) and material-induced electrostrictive forces (photo-elastic effects), in a manner consistent with Refs. \cite{rakich2012giant, qiu2013stimulated, PhysRevA.92.013836, laude2015lagrangian}.  Hence, this treatment is directly applicable to a range of emerging micro- and nano-scale optomechanical systems \cite{kang2009tightly, shin2013tailorable, van2015interaction, kittlaus2015large, shin2015control, dainese2006raman, kang2010all}.  Based on this self-consistent fully coupled nonlinear treatment, we establish the connection between the power spectral density of spontaneously scattered light in forward Brillouin interactions and the nonlinear coupling strength, which is expressed both in terms of an optomechanical coupling rate and the more conventional Brillouin gain coefficient. 

In sharp contrast to backward Brillouin scattering, we show that, noise-initiated stimulated forward Brillouin scattering is forbidden in the majority of experimental systems since the anti-Stokes and Stokes waves interact through the same phonon mode. In fact, our analysis shows that the single-pass gain, which conventionally characterizes the threshold for energy transfer in back-scattering processes, is negative for a large class of forward Brillouin devices.  Interestingly, the spontaneous noise grows linearly whereas the signal amplifies quadratically with device length in the weak signal limit. However, in waveguides with high optical dispersion or in inter-modal scattering, distinct phonon modes mediate Stokes and anti-Stokes scattering. This dispersive symmetry breaking leads to exponential optical amplification and noise dynamics that are reminiscent of backward Brillouin scattering.

\section{\label{sec:level1} Theoretical Study}
\begin{figure}[tbh]
\centerline{
\includegraphics[width=9.0cm]{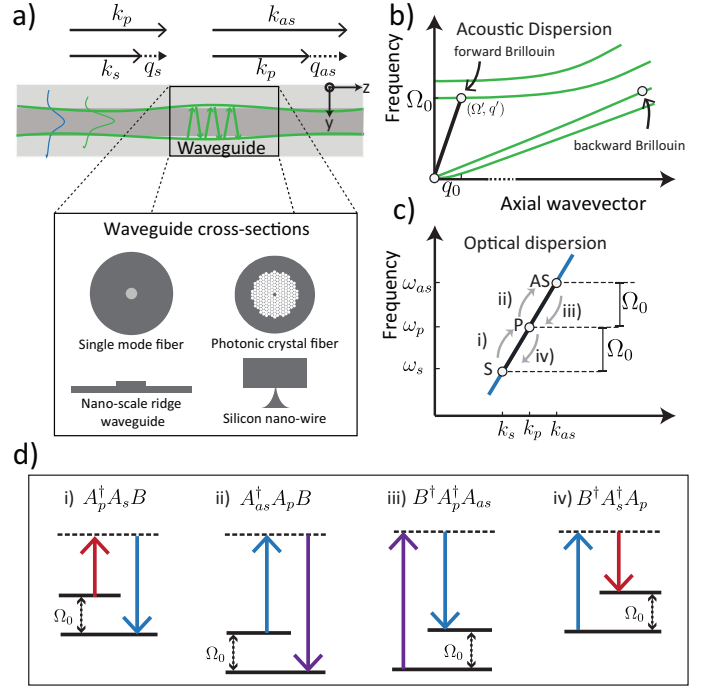}}
\caption{(a) A general schematic of a waveguide that supports both acoustic and optical modes. Different example waveguide geometries include:  Single mode fiber \cite{shelby1985guided}, photonic crystal fiber \cite{kang2009tightly}, nano-scale ridge waveguides \cite{kittlaus2015large} and silicon nano-wire \cite{van2015interaction} (b) Dispersion curves for acoustic modes inside a waveguide. The acoustic modes relevant to forward Brillouin are optical-phonon-like modes  with cut-off frequency, $\Omega_0$. (c) Phase matching requirements allow each phonon mode to interact with a set of higher order Stokes and anti-Stokes photon modes. (d) Set of Brillouin interactions that underlie complex dynamics where both photons and phonons are coherently created and annihilated.} 
\label{fig1}
\end{figure}

We consider interactions between co-linearly propagating optical- and elastic-waves within a general class of optomechanical waveguides that support guidance of both photons and phonons; example systems are shown schematically in Fig. \ref{fig1}(a).  This translationally invariant waveguide, has a transverse profile specified by $\epsilon_r(\bf{r_{\perp}})$, $\rho(\bf{r_{\perp}})$, and $c_{ijkl}(\bf{r_{\perp}})$, representing the dielectric distribution, mass density distribution, and elastic tensor profile, respectively. 
We express the guided modes of the system in terms of electric and acoustic displacement fields  $\text{\bf{D}}_l({\bf r},t) = \text{\bf{d}}_l(\textbf{r}_\perp) e^{i(k_l z-\omega_l t)}$ and $\text{\bf{U}}_m({\bf r},t) = \boldsymbol{\mathscr{U}}_m(\textbf{r}_\perp) e^{i(q_m z-\Omega_m t)}$ respectively. These modes are obtained by solving Maxwell's equations and the elastic wave equation, $\partial_jc_{ijkl}\partial_k \text{U}_{l,m} = -\Omega_m^2 \rho \text{U}_{i,m}$ \cite{royer2000elastic}. Here, $\text{\bf{d}}_l(\textbf{r}_\perp)$ is the electric displacement profile of an optical mode with wave-vector, $k_l$, and frequency $\omega_l $. Similarly, $\boldsymbol{\mathscr{U}}_m(\textbf{r}_\perp)$ is the elastic displacement profile of a phonon mode with wave-vector, $q_m$, and frequency $\Omega_m$. Note that the set of points $\{\omega_l,k_l\}$ and $\{\Omega_m,q_m\}$ lie on the optical dispersion curves, $\omega(k)$ and $\Omega(q)$, as seen in Fig. \ref{fig1}(b) and (c), permitting alternative representations $\{\omega(k_l),k_l\}$ and $\{\Omega(q_m),q_m\}$.

In what follows, we consider noise initiated scattering of energy from an incident monochromatic pump wave $(\textbf{d}_{p}(\textbf{r}_\perp), k_p, \omega_p)$ by one or more Brillouin active phonon modes. A characteristic Brillouin-active phonon mode is denoted by $(\Omega',q')$ on the dispersion curve in Fig. \ref{fig1}(b). We begin by considering intra-modal scattering, or coupling between optical waves of distinct frequency that are guided in the same optical band (Fig. \ref{fig1}(c)). 
Pump photons can be red-shifted to mode $(\textbf{d}_{s}(\textbf{r}_\perp), k_s, \omega_s)$ through a Stokes process,  or blue-shifted to mode $(\textbf{d}_{as}(\textbf{r}_\perp), k_{as}, \omega_{as})$ through anti-Stokes processes. Note that phase-matching (Fig. \ref{fig1}(a)) and energy conservation (Fig. \ref{fig1}(d)) yield distinct requirements for Stokes and anti-Stokes processes:\begin{align} \label{stokesPhaseMatchin}
& \Omega(q_s)\,\,= \omega(k_p)-\omega(k_s), ~ \quad  q_s = k_p-k_s, \\ \label{antistokesPhaseMatchin}
& \Omega(q_{as}) = \omega(k_{as})-\omega(k_p), \quad  q_{as} = k_{as}-k_p.
\end{align}
Hence, $\Omega_s\neq\Omega_{as}$ and $q_{s} \neq q_{as}$, meaning that, in general, the Stokes phonon $(\Omega',q')\rightarrow (\Omega_s, q_s)$ and the anti-Stokes phonon $(\Omega',q')\rightarrow (\Omega_{as}, q_{as})$ are non-degenerate. In other words, for an ideal system of infinite extent, the Stokes and anti-Stokes processes are mediated by independent phonon modes. In the following sections, we will see that this form of `dispersive' symmetry breaking between Stokes and anti-Stokes processes strongly impacts the system dynamics.  However, in many practical (or finite) systems, dispersive symmetry breaking becomes a subtle consideration.

The distinct Stokes and anti-Stokes phonon modes (defined above) are not resolved through intra-modal coupling in numerous forward Brillouin systems \cite{kang2009tightly, shin2013tailorable, van2015interaction, kittlaus2015large, shin2015control, dainese2006raman}; hence, the Stokes and anti-Stokes scattering processes effectively couple to the same phonon mode.  To understand why, we begin by Taylor expanding $\omega(k)$ in Eqs. (\ref{stokesPhaseMatchin}) and (\ref{antistokesPhaseMatchin}) to find $q_{s} \cong \Omega_s/v_g(k_p)$ and $q_{as} \cong \Omega_{as}/v_g(k_{as})$, where $v_g(k) \equiv (\partial \omega/\partial k)_{k} $. These expressions reveal that  $q_{s}$ and $q_{as}$ are very small, pushing $\Omega_{s}$ and $\Omega_{as}$ very near the phonon cutoff frequency $\Omega_{o}$, as seen in Fig. \ref{fig1}(b).  With $\Omega_{as}\cong \Omega_{s} \cong \Omega_{o}$, one finds $\Delta q = |q_{as}\!-\!q_{s}| \cong (\partial^2 k/\partial \omega^2)\Omega^2_o$. Therefore, in the case when the propagation length is much less than $\pi/\Delta q$, wave uncertainty reveals that the Stokes and anti-Stokes phonons are not resolvable; to an excellent approximation both optical processes couple to the phonon state $(\Omega_{o}, q_{o})$ where $q_{o} \equiv \Omega_o/v_g(k_p)$. As a result, we will see that the equations of motion that govern Stokes and anti-Stokes generation are intimately coupled. 

By contrast, distinct Stokes $(\Omega_{s}, q_{s})$ and anti-Stokes $(\Omega_{as}, q_{as})$ phonons are well resolved through more widely studied backwards Brillouin coupling (e.g., see $(\Omega_{s}, q_{s})$ in Fig. \ref{fig1}(b)). This is because the scattered Stokes and anti-Stokes waves propagate contra-directionally to the pump wave in the backward case. Solving Eqs. \ref{stokesPhaseMatchin} in the case of contra-directional coupling, one finds, $q_s = k_p-k_s \approx 2|k_p|$ and $q_{as} = k_{as}-k_p \approx -2|k_p|$ \cite{boyd2003nonlinear, agrawal2007nonlinear}. Since $\Delta q$ is large ($\sim 4 |k_p|$), the Stokes and anti-Stokes phonon modes are very well resolved through backwards Brillouin interactions, resulting in independent equations of motion for Stokes and anti-Stokes generation. In what follows, we show that noise properties and nonlinear dynamics of forward Brillouin processes differ sharply from the more widely studied backward Brillouin processes \cite{boyd1990noise, gaeta1991stochastic, yeniay2002spontaneous, le2003theoretical}; this distinct behavior hinges on the absence or presence of dispersive symmetry breaking.

In what follows, we begin by applying the general Hamiltonian framework of Section \ref{sec:level2} to this frequent case (i.e., coupling to same phonon mode) in sections \ref{sec:level3}-\ref{sec:level5}.  In section \ref{sec:level7} we return to the cases when the Stokes and anti-Stokes phonon degeneracies are resolvable through forms of dispersive symmetry breaking.
 
%
%
%

\subsection{\label{sec:level2}Hamiltonian of a forward Brillouin system}
Building on the quantum traveling-wave treatment of Brillouin coupling by Sipe {\it et al.} \cite{sipe2015hamiltonian}, and prior quantum-traveling wave treatments of noise and nonlinearity \cite{haus1970steady, imoto1985quantum, blow1990continuum, jeffers1993quantum, haus2000electromagnetic, loudon2000quantum}, we present a Hamiltonian treatment captures the distributed optomechanical coupling, noise, and spatio-temporal field evolution in Brillouin interactions. This formulation incorporates the optomechanical couplings produced by boundary-induced radiation pressures (boundary motion) and material-induced electrostrictive forces (photo-elastic effects), in a manner consistent with Refs. \cite{rakich2012giant, qiu2013stimulated, PhysRevA.92.013836, laude2015lagrangian}. Hence, this treatment is directly applicable to a range of emerging micro- and nano-scale optomechanical systems \cite{kang2009tightly, shin2013tailorable, van2015interaction, kittlaus2015large, shin2015control, dainese2006raman, kang2010all}.  

The Hamiltonian for forward Brillouin scattering is written as
\begin{align}
H = H^{\text{ph}}+H^{\text{opt}}+H^{\text{int}}.
\end{align}
Here $H^{\text{ph}}$, $H^{\text{opt}}$, and $H^{\text{int}}$
characterize; the dynamics of the acoustic field, the dynamics of the optical field, and  
the acousto-optic interaction. 
For a waveguide that is axially invariant in $z$, the acoustic Hamiltonian can be written as follows
\begin{align} \label{Hph0}
H^{\text{ph}} = \int\text{d}q \ \hbar \Omega(q) b_q^{\dagger} b_q,
\end{align}
Here $b_q$ is the annihilation operator for the $q$th phonon mode and we are summing over all the acoustic wavevectors to allow for spatial evolution of the acoustic field along the waveguide. We then introduce phonon mode envelope operator $B(z) = 1/\sqrt{2\pi}\int \text{d}q \ b_q e^{i(q-q_0)z}$, with carrier wavevector $q_0$. Taking Fourier transform to express $b_q$ in terms of $B(z)$, we write down a Hamiltonian that emphasizes spatial evolution of fields in Brillouin scattering
\begin{align} \label{Hph}
H^{\text{ph}} = \int\text{d}q \ \hbar \Omega(q) b_q^{\dagger} b_q = \int\text{d}z \ \hbar B^{\dagger}(z) \hat{\Omega}_z B(z).
\end{align}
The spatial operator $\hat{\Omega}_z$, resulting from the group velocity dispersion of the phonon frequency $\Omega(q)$, is given by 
\begin{align}\label{OzoperatorPhonon}
\hat{\Omega}_z=\sum_{n=0}^{\infty} \frac{1}{n!} \frac{\partial^n \Omega}{\partial q^n}\biggr\rvert_{q_0}\left(-i \frac{\partial}{\partial z}\right)^n.
\end{align}
For the slow group velocity guided acoustic mode relevant to forward Brillouin scattering, $\hat{\Omega}_z \approx \Omega_0-i {\rm v}_0 \frac{\partial}{\partial z},$ where ${\rm v}_0$ is the acoustic group velocity; the slowly-varying envelope approximation (SVEA) is obtained by
ignoring higher order dispersion terms in (\ref{OzoperatorPhonon}). 
The acoustic Hamiltonian in equation (\ref{Hph}) within the SVEA is given by 
\begin{align}
H^{\text{ph}} = \int{\text{d}z \left( \hbar \Omega_0 B^{\dagger}(z)B(z)-i\hbar {\rm v_0} B^{\dagger}(z)\partial_z B(z) \right).}
\end{align}

The \emph{traveling} nature of phonons in Brillouin scattering and group velocity dispersion of the acoustic modes makes the acoustic Hamiltonian different than that for a \emph{stationary} phonon mode, $\hbar \Omega_0 b^{\dagger} b,$ in cavity optomechanics. The term $\text{v}_0B^{\dagger}(z)\partial_z B(z)$ captures the traveling nature of the phonons, leading to a spatial change in the acoustic field along the waveguide. Therefore, $B^{\dagger}(z)B(z)$ is reminiscent of the phonon number density operator. This is somewhat analogous to the phonon number operator in cavity optomechanics $b^{\dagger}b$. The acoustic power flow in the waveguide is given in terms of the envelope operator as \cite{sipe2015hamiltonian}
\begin{align}
P_{\text{ph}} &= \hbar \Omega_0 \text{v}_0 B^{\dagger}(z,t)B(z,t). 
\label{powac}
\end{align}
A similar analysis gives the Hamiltonian for the optical fields within the SVEA as
\begin{multline}
H^{\text{opt}} = \sum_{\gamma} \int \text{d}k \ \hbar \omega_{\gamma}(k)a_{\gamma k}^{\dagger} a_{\gamma k}  \\
=\sum_{\gamma} \int{\text{d}z   \left ( \hbar \omega_{\gamma}A^{\dagger}_{\gamma}(z)A_{\gamma}(z) - i\hbar v_{\gamma}A^{\dagger}_{\gamma}(z) \partial_z A_{\gamma}(z)  \right)},
\end{multline}
where $A_{\gamma}(z) = 1/\sqrt{2\pi}\int \text{d}k \ a_{\gamma k} e^{i(k-k_{\gamma})z}$ is the optical mode envelope operator and  $v_\gamma$ is the optical group velocity for the anti-Stokes, pump and Stokes field (i.e. $\gamma = \text{as, p and s}$). The optical power flow, $P_{\gamma}$, is given by
\begin{align} \label{powopt}
P_{\gamma} &= \hbar \omega_{\gamma} v_{\gamma} A^{\dagger}_{\gamma}(z,t)A_{\gamma}(z,t).
\end{align}
Finally, the acoustic-optic coupling is captured in the interaction Hamiltonian given by (see Appendix)
\begin{multline} \label{Hint}
H^{\text{int}} = \hbar \int{ \text{d}z}  \Bigl(  g_0A^{\dagger}_p(z)A^{}_s(z) B(z) \\  
 +  g_1A^{\dagger}_{as}(z)A^{}_p(z) B(z) \Bigr) 
+ {\rm H.c.}
\end{multline}
where ${\rm H.c.}$ stands for Hermitian conjugate. The coupling rates $g_0 = g_{\text{p,s}}$ and $g_1= g_{\text{as,p}}$) describe two forward Brillouin processes: the annihilation of a Stokes photon and a phonon to  create a pump photon (i.e. $A_p^{\dagger}A_sB$), and the annihilation of pump photon and a phonon to create an anti-Stokes photon (i.e. $A_{as}^{\dagger}A_pB$). The coupling rates for conjugate processes are given by $g_0^*$ and $g_1^*$. 
These coupling rates are given by the following overlap integral
\begin{align}
& g_{\gamma,\gamma^{\prime}} \! = \!
 \frac{1}{\epsilon_o}\sqrt{\frac{\omega_{\gamma}}{2}} \sqrt{\frac{\omega_{\gamma^{\prime}}}{2}} \sqrt{\frac{\hbar \Omega_{0}}{2}} \int{(d_{\gamma}^i(\textbf{r}_\perp))^*d_{\gamma^{\prime}}^j(\textbf{r}_\perp)}  \\
& \times \left( p^{ijlm}(\textbf{r}_\perp) \frac{\partial \mathscr{U}_{0}^l (\textbf{r}_\perp)}{\partial r^m}-\delta^{ij} \left(\frac{\partial \beta_{r}(\textbf{r}_\perp)}{\partial r^l} \right)\mathscr{U}_{0}^l (\textbf{r}_\perp)\right)d\textbf{r}_{\perp}. \nonumber 
\end{align}
The acoustic modes for a medium with photoelastic constants $p^{ijlm}$, density $\rho$, and relative dielectric index $ \epsilon_r=1/\beta_r,$ are normalized such that $\Omega^2_{0} \int{\text{d}\textbf{r}_{\perp} \rho(\textbf{r}_{\perp})\boldsymbol{\mathscr{U}}^*_{0}(\textbf{r}_\perp)\cdot \boldsymbol{\mathscr{U}}_{0}(\textbf{r}_\perp)} = 1$ and $\frac{1}{\epsilon_o} \int{\text{d}\textbf{r}_{\perp} \beta_r(\textbf{r}_{\perp}) \textbf{d}^*_{\gamma}(\textbf{r}_\perp) \cdot \textbf{d}_{\gamma}(\textbf{r}_\perp)} = 1$.
These coupling terms, which are essentially transverse mode overlaps, encapsulate both the photo-elastic interaction and the effect of radiation pressure and moving boundaries \cite{sipe2015hamiltonian}. Importantly, the full-vectorial nature of the electric and acoustic displacement fields allows calculation of coupling rates for both intra- and inter-modal coupling.

Generalizing the Hamiltonian to account for the multi-wave parametric interaction that leads to novel dynamics, such as frequency comb generation, is done by including all higher order Stokes and anti-Stokes fields in $H^{\text{opt}}$. Similarly, the interaction Hamiltonian, $H^{\text{int}},$ must be expanded to include all possible interactions that lead to creation or annihilation of phonons as follows
\begin{multline}
H^{\text{int}} = 
\int dz \left( \sum_{n}  \hbar g_n A^{\dagger}_n(z)A^{}_{n-1}(z) B(z) +{\rm H.c.} \right )
\end{multline}
where the integer $n$ indexes the pump field at frequency $\omega_{n=0}$ and  all higher order Stokes and anti-Stokes field at frequency $\omega_n = \omega_0+n\Omega_0$. For the rest of this paper, we consider the simplified interaction Hamiltonian in equation (\ref{Hint}).
\subsection{\label{sec:level3} Equations of motion}
Using this Hamiltonian, the spatio-temporal evolution of the envelope fields is obtained from the Heisenberg equations of motion (see Appendix A)
\begin{align}\label{hisenbergeq}
\frac{\partial B(z,t)}{\partial t} &= \frac{1}{i\hbar}[B(z,t),H], \\ \label{hisenbergeq2}
\frac{\partial A_{\gamma}(z,t)}{\partial t} &= \frac{1}{i\hbar}[A_{\gamma}(z,t),H].
\end{align}
The equations above are evaluated by using the following equal time commutation relations for the envelope field operators (see Appendix A)
\begin{align}
[B_{\alpha}(z,t),B^{\dagger}_{\alpha^{\prime}}(z^{\prime},t)] &= \delta_{\alpha \alpha^{\prime}} \delta(z-z^{\prime}), \\
[A_{\gamma}(z,t),A^{\dagger}_{\gamma^{\prime}}(z^{\prime},t)] &= \delta_{\gamma \gamma^{\prime}} \delta(z-z^{\prime}).   
\end{align}
After factoring out the fast oscillating component of the envelope field operators by letting $\bar{B}(z,t) = B(z,t)e^{i\Omega t}$, where $\Omega = \omega_p-\omega_s$ is the detuning between the pump and the Stokes light, and $\bar{A}_{\gamma}(z,t) = A_{\gamma}(z,t)e^{i\omega_\gamma t}$, Eqs. (\ref{hisenbergeq}) and (\ref{hisenbergeq2}) give the following spatio-temporal evolution
\begin{align} 
\label{phonEq1}
& \frac{\partial \bar{B}}{\partial t} +\text{v}_0 \frac{\partial \bar{B}}{\partial z} \! = i(\Omega\!-\!\Omega_0)\bar{B}\!-\!i \left( g_0^* \bar{A}^{\dagger}_s \bar{A}_p \!+ \!g_1^* \bar{A}^{\dagger}_p \bar{A}_{as}\right)
\\ \label{pumpEq}
& \frac{\partial \bar{A}_p}{\partial t} + v_p \frac{\partial \bar{A}_p}{\partial z} = -i \left( g_0 \bar{A}^{}_s \bar{B}\!+\! g_1^* \bar{B}^{\dagger} \bar{A}_{as}\right) 
\\ \label{stokesEvEq}
&\frac{\partial \bar{A}_s}{\partial t} + v_s \frac{\partial \bar{A}_s}{\partial z} = -i g_0^* \bar{B}^{\dagger} \bar{A}_p\\ \label{antistokesEq}
&
\frac{\partial \bar{A}_{as}}{\partial t} + v_{as} \frac{\partial \bar{A}_{as}}{\partial z} = -ig_1 \bar{A}_p \bar{B}.
\end{align}
These equations of motion are similar to the ones derived classically using nonlinear polarization and density variation induced by electrostrictive forces \cite{kang2009tightly,kang2010all}. In addition, the coupling term here accounts for both electrostrictive and radiation pressure forces, extending its validity to nano-scale systems.

To capture the quantum statistical physics of the phonon field we introduce a dissipation rate, $\Gamma_0/2,$ and a Langevin force, $\eta(z,t),$ arising due to thermal fluctuations. For most forward Brillouin processes of our interest the dissipation rate is large (i.e. in the MHz range) and the group velocity is vanishingly small ($\sim$ 1 m/s) \cite{kang2009tightly}. Therefore, we ignore the $\partial B/\partial z$ term representing the spatial evolution of envelope field in equation (\ref{phonEq1}). In this case, the phonon mode amplitude satisfies the following equation of motion
\begin{multline} \label{phononEq}
\frac{\partial \bar{B}}{\partial t} \!=\! 
i(\Omega-\Omega_0)\bar{B} -\frac{\Gamma_0}{2}\bar{B}-i \left( g_0^* \bar{A}^{\dagger}_s \bar{A}_p \!+\! g_1^* \bar{A}^{\dagger}_p \bar{A}_{as}\right)+\eta.
\end{multline}
We assume that dissipation and thermal fluctuations of the optical fields is negligible; 
at typical lab temperatures the thermal occupancy of optical photons is very small. In section (\ref{sec:level5}) we show that thermal fluctuations of the phonon mode lead to spontaneous scattering of light. However, before exploring the spontaneous forward Brillouin scattering (forward spontaneous noise), we study the stimulated regime. 

In the next section we derive the gain coefficient, $G_B$(W$^{-1}$m$^{-1}$), for stimulated forward Brillouin scattering in terms of the coupling rates of the Hamiltonian. Eventually, we relate this Brillouin gain coefficient to the spontaneous forward scattering efficiency, allowing stimulated measurements to predict spontaneous scattering.

\subsection{\label{sec:level4} Stimulated forward Brillouin scattering}
In the presence of a number of driven optical fields the steady state phonon envelope field in equation (\ref{phononEq}) reduces to 
\begin{align} \label{phononsteadyState}
\bar{B}(z,t) = \frac{\left( g_0^* \bar{A}^{\dagger}_s\bar{A}_p + g_1^* \bar{A}^{\dagger}_p \bar{A}_{as}\right)}{\left( \Omega-\Omega_0+i\frac{\Gamma_0}{2}\right)}
\end{align}
where we've assumed $\omega_{as} - \omega_p = \omega_p- \omega_s = \Omega$, and  
where a negligible contribution from the Langevin force has been dropped. Substituting Eq. (\ref{phononsteadyState}) into  Eq. (\ref{stokesEvEq}) we get the following steady state spatial evolution for the Stokes field
\begin{align} \label{stokessteadyState}
\frac{\partial \bar{A}_s}{\partial z} &= -\frac{i}{v_s}\frac{g_0^*(g_0 \bar{A}^{\dagger}_p\bar{A}_s + g_1 \bar{A}^{\dagger}_{as} \bar{A}_{p}) \bar{A}_p}{\left( \Omega-\Omega_0-i\frac{\Gamma_0}{2}\right)}.
\end{align}
This solution takes into account the back-action of the phonon-field. Before solving for the Stokes field amplitude, we consider the weak signal limit to define Brillouin gain coefficient. In the undepleted pump regime ($|A_{p}| \gg |A_s|$ and $|A_{as}|$) we define Brillouin gain coefficient as: $dP_s/dz = G_B P_p P_s$, where $G_B$ is the Brillouin gain coefficient, $P_s$ and $P_p$ are the powers in the Stokes and the pump field respectively. Assuming $A_{as} \rightarrow 0$, the first term on the right hand side of equation (\ref{stokessteadyState}) and the expression for power in the optical fields given by equation (\ref{powopt}), gives $G_B$ in terms of the coupling strength as follows
\begin{align}\label{bg}
G_B = \frac{4|g_0|^2}{v_s v_p \Gamma_0 \hbar \omega_p} \frac{\left(\frac{\Gamma_0}{2}\right)^2}{(\Omega-\Omega_0)^2+(\frac{\Gamma_0}{2})^2}.
\end{align}

The equations of motion, accounting for the coupled dynamics of the Stokes and anti-Stokes fields in the undepleted pump regime, give the following steady state Stokes amplitude at position $z=L$
\begin{align}
\left\langle |\bar{A}_s(L)|^2 \right\rangle & \approx |\bar{A}_s(0)|^2 \left(1+\frac{2|\bar{A}_p|^2 |g_0|^2 L}{v\Gamma_0}\right)^2
\end{align}
where the single pass gain is assumed small (to be discussed in the next section Eq. (\ref{singlePassgain})).
Above, $\bar{A}_s(0)$ and $\bar{A}_p$ are the input Stokes and pump field, and we have assumed that the input anti-Stokes field is zero, i.e. $\bar{A}_{as}(0)=0.$ Therefore, for small single pass gain, the Stokes power for stimulated forward Brillouin grows algebraically with length.

\subsection{\label{sec:level5} Spontaneous forward Brillouin scattering}
In this section, we derive spontaneous scattering of pump light into co-linearly propagating Stokes and anti-Stokes fields that result from thermally driven guided acoustic modes. Before solving the coupled equations (\ref{pumpEq}-\ref{phononEq}), we explore the statistical properties of the Langevin force, $\eta (z,t),$ by using the distributed, fluctuating source model first presented by Boyd \emph{et al.} to describe spontaneous backward Brillouin scattering \cite{boyd1990noise}. 

\subsubsection{\label{subsec:level6} Properties of the Langevin force}
For conceptual development we divide the waveguide into small subregions of length $\Delta z$ such that $\bar{B}$ is effectively constant in the subregion. Let $\bar{B}_i$ and $\eta_i$ denote the acoustic envelope field and the Langevin force averaged over the $i^{\text{th}}$ subregion. Then, $\bar{B}_i^{\dagger}\bar{B}_i$ represents the phonon density operator for the $i^{\text{th}}$ subregion. We assume that $\eta_i$ is a Gaussian random variable with the following properties
\begin{align} \label{discreteLang}
\left\langle \eta_i \right\rangle = 0, \text{ and } \left\langle \eta_i^{\dagger}(t) \eta_j(t^{\prime}) \right\rangle = \tilde{Q} \delta_{ij} \delta(t-t^{\prime}).
\end{align}
Here, $\tilde{Q}$ characterizes the strength of the fluctuations in $\eta_i$. To find $\tilde{Q}$ we relate fluctuations in $\bar{B}_i$ to the fluctuation of $\eta_i$, and demand that $\bar{B}_i^\dag\bar{B}_i$ is given by the thermal number density of a phonon mode of frequency $\Omega_0$ in equilibrium. Without driving due to optical forces $\bar{B}_i$ obeys the following equation
\begin{equation}
\frac{d\bar{B}_i}{dt} = -\frac{\Gamma_0}{2}\bar{B}_i+\eta_i.
\end{equation}
With the solution $\bar{B}_i(t) = \int_{-\infty}^t dt^{\prime} e^{-\Gamma_0(t-t^{\prime})/2} \eta_i(t^{\prime})$ we get the following equal time correlation
\begin{equation} \label{phcor}
\left\langle \bar{B}_i^{\dagger}(t)\bar{B}_j(t) \right\rangle = \delta_{ij} \frac{\tilde{Q}}{\Gamma_0}.
\end{equation}
We now find $\tilde{Q}$ by requiring that average phonon density for the thermally driven mode is given by
\begin{align} \label{phcortherm}
\left\langle \bar{B}_i^{\dagger}(t)\bar{B}_i(t) \right\rangle = \frac{\bar{n}_{\text{th}}}{\Delta z},
\end{align}
where $\bar{n}_{\text{th}} = 1/(e^{\hbar \Omega_0/k_\text{B}T}-1)$ is the average number of thermal phonons of angular frequency $\Omega_0$ at temperature $T$. 
Using equation (\ref{phcor}) and (\ref{phcortherm}) we have
\begin{align}
\tilde{Q} = \frac{\bar{n}_{\text{th}} \Gamma_0}{\Delta z}.
\end{align}
Finally, taking the continuum limit of equation (\ref{discreteLang}) we find
\begin{align} 
\left\langle \eta(z,t) \right\rangle &= 0, \\
\left\langle \eta^{\dagger}(z,t) \eta(z^{\prime},t^{\prime}) \right\rangle &= Q \delta(z-z^{\prime})\delta(t-t^{\prime}),\label{contLangevin}
\end{align}
where the strength of fluctuation, $Q$, is given by
\begin{align} \label{StrengthOfFluctuation}
Q = \tilde{Q} \Delta z = \bar{n}_{\text{th}}\Gamma_0. 
\end{align}
It is important to mention that because of the commutation relation for the phonon mode amplitude operator (i.e. $[B(z,t),B^{\dagger}(z^{\prime},t)]=\delta(z-z^{\prime})$),
\begin{align}\label{contLangevin2}
\left\langle \eta(z,t) \eta^{\dagger}(z^{\prime},t^{\prime}) \right\rangle = (\bar{n}_{\text{th}}+1)\Gamma_0 \delta(z-z^{\prime})\delta(t-t^{\prime}).
\end{align}
However, in the high temperature limit (i.e. classical limit), $\bar{n}_{\text{th}} \simeq k_{\text{B}} T/\hbar \Omega_0 \gg 1$, meaning  $\left\langle \eta\eta^{\dagger}\right\rangle \simeq  \left\langle \eta^{\dagger} \eta \right\rangle.$

\subsubsection{\label{subsec:level6} Spontaneous forward scattering efficiency}
To compute spontaneous forward Brillouin scattering, we assume an undepleted pump and no input Stokes or anti-Stokes field in the waveguide. We solve the coupled mode equations (\ref{stokesEvEq}-\ref{phononEq}), assuming that the group velocity for Stokes and anti-Stokes light are the same. This calculation gives the following solution for the Stokes and the anti-Stokes envelope fields \cite{boyd1990noise, raymer1981stimulated}
\begin{widetext}
\begin{align}\label{stokesComplete}
&\bar{A}_s(z,\tau) = -i\frac{g_0^*}{v} |\bar{A}_p| \int_0^{\tau}{\text{d}\tau^{\prime}\int_0^{z}{\text{d}z^{\prime} \  \eta^{\dagger}(z^{\prime},\tau^{\prime})e^{-\frac{\Gamma_0}{2}(\tau-\tau^{\prime})} \text{I}_0\left( \left[\frac{4}{v}\left(|g_0|^2-|g_1|^2\right)|\bar{A}_p|^2(\tau-\tau^{\prime})(z-z^{\prime}) \right]^{1/2} \right)}}, \\ \label{antiStokesComplete}
&\bar{A}_{as}(z,\tau) = -i\frac{g_1}{v} |\bar{A}_p| \int_0^{\tau}{\text{d}\tau^{\prime}\int_0^{z}{\text{d}z^{\prime} \  \eta^{}(z^{\prime},\tau^{\prime})e^{-\frac{\Gamma_0}{2}(\tau-\tau^{\prime})} \text{I}_0\left( \left[\frac{4}{v}\left(|g_0|^2-|g_1|^2\right)|\bar{A}_p|^2(\tau-\tau^{\prime})(z-z^{\prime}) \right]^{1/2} \right)}}. 
\end{align}
\end{widetext}
Here, I$_n(x)$ is the modified Bessel of the first kind and we have switched the co-ordinate system from $(z,t)$ to the retarded frame $(z,\tau=t-z/v)$. From these expressions and using the statistical properties of the Langevin force derived in (\ref{contLangevin}) and (\ref{contLangevin2}),  we get the following spontaneously scattered stokes and anti-Stokes signal in the long time limit $\tau \rightarrow \infty$ at position $z=L$
\begin{align}
\resizebox{1.0\hsize}{!}{$\left\langle |\bar{A}_s(L)|^2 \right\rangle = \frac{|g_0|^2}{v^2} |\bar{A}_p|^2 L(\bar{n}_{\text{th}}+1)e^{\frac{G}{2}} \left( I_0\left(G/2\right)- I_1\left(G/2\right) \right)$},
\end{align}
\begin{equation}
\resizebox{0.9\hsize}{!}{$\left\langle |\bar{A}_{as}(L)|^2 \right\rangle = \frac{|g_1|^2}{v^2} |\bar{A}_p|^2 L \bar{n}_{\text{th}} e^{\frac{G}{2}} \left( I_0\left(G/2\right)- I_1\left(G/2\right) \right)$},
\end{equation}
where  
\begin{align} \label{singlePassgain}
G = \frac{4}{v}\left(|g_0|^2-|g_1|^2\right) \frac{L |\bar{A}_p|^2}{\Gamma_0}
\end{align}
is the single pass gain for forward Brillouin scattering. Note that this $G$ is not to be confused with the Brillouin gain coefficient $G_\text{B}$ derived earlier. Assuming that the mode profiles are the same for the Stokes and anti-Stokes fields we can show that  
\begin{align}
G = - \frac{2\Omega_0}{\omega_s} G_{\text{B}}P_p L .
\end{align}
The ratio $\Omega_0/\omega_s$ is typically of the order of $10^{-5}$ for a phonon mode in the GHz range and a photon mode in the 200 THz range. In contrast to backward spontaneous Brillouin scattering \cite{boyd1990noise}, $G,$ is negative and close to zero for forward spontaneous scattering. This result is consistent with the fact that for most forward Brillouin interactions there is no symmetry breaking between the Stokes and the anti-Stokes processes (unless there is a strong optical dispersion); phonons created in Stokes scattering are annihilated in anti-Stokes processes. 
Since $G \approx 0$, the forward Brillouin scattering efficiency, $E_F$, which is defined as the ratio of total power generated in the Stokes or anti-Stokes fields at position $L$ along the waveguide to the input pump light power, is given by
\begin{align}
E_{F,s} &= \frac{\left\langle |\bar{A}_s(L)|^2 \right\rangle}{|\bar{A}_p|^2}= \frac{|g_0|^2}{v^2}(\bar{n}_{\text{th}}+1)L \\ 
E_{F,as} &= \frac{\left\langle |\bar{A}_{as}(L)|^2 \right\rangle}{|\bar{A}_p|^2}= \frac{|g_1|^2}{v^2}\bar{n}_{\text{th}}L.
\end{align}

Therefore, unlike backward stimulated Brillouin scattering \cite{boyd1990noise}, noise does not grow exponentially for forward Brillouin scattering. In the high temperature limit, the scattering efficiency can be written in terms of the peak Brillouin gain coefficient ($G_B(\Omega_0)$) derived in equation (\ref{bg}) and is given by
\begin{align} \label{stimtospont}
E_{F,as} \simeq E_{F,s} = \frac{\omega_p G_B k_{\text{B}} T L \Gamma_0}{4\Omega_0}.
\end{align}
This expression above relates Brillouin gain coefficient, $G_\text{B},$ which can be measured from stimulated forward light scattering measurements, to the light spontaneously scattered in the forward direction by thermally excited guided acoustic modes. 
\subsubsection{Stokes field correlation function and power spectrum}
In the limit of large acoustic damping 
a simple form for the temporal and spatial dependent correlations of the Stokes field can be derived. In this limit the phonon envelope is determined by its instantaneous steady-state value given by 
\begin{equation}
\label{ }
\bar{B}(z,t) \approx \bar{B}_{th}(z,t) - \frac{i \left( g_0^* \bar{A}^{\dagger}_s \bar{A}_p \!+\! g_1^* \bar{A}^{\dagger}_p \bar{A}_{as}\right)}{i\Delta+\frac{\Gamma_0}{2}} 
\end{equation}
where $\Delta = \Omega_0-\Omega$ and 
\begin{equation}
\label{ }
\bar{B}_{th}(z,t) = \int_{-\infty}^t d\tau \ e^{-\left(i \Delta +\frac{\Gamma_0}{2}\right)(t-\tau)}\eta(z,\tau).
\end{equation}
This solution for $\bar{B}(z,t)$ can now be directly substituted into the equations of motion for the Stokes and anti-Stokes envelopes to give
\[
\begin{bmatrix}
    \hat{L}_s    & -\chi g_0^*g_1 \bar{A}_p^2   \\
    \chi g_0 g_1^* \bar{A}{_p^\dag}^2     & \hat{L}_{as}   \\
\end{bmatrix}
\begin{bmatrix}
    \bar{A}_s(z,t)  \\
    \bar{A}_{as}^\dag(z,t)  \\
\end{bmatrix}
= -i
\begin{bmatrix}
    g_0^* \bar{A}_p\\
    -g_1^*\bar{A}_p^\dag  \\
\end{bmatrix} \bar{B}^\dag_{th}(z,t)
\]
where 
\begin{align}
\label{}
     \chi & = [-i\Delta+\Gamma_0/2]^{-1}  \\
     \hat{L}_s & = \partial_t+v_s \partial_z \! -\!  \chi |g_0|^2 |\bar{A}_p|^2 \\
     \hat{L}_{as} & = \partial_t+v_{as} \partial_z \! +\! \chi |g_1|^2 |\bar{A}_p|^2.
\end{align}
We assumed undepleted pump and $v_s \approx v_{as} = v$. 
These coupled equations can be manipulated to give the equation of motion for the Stokes field, including the back reaction from anti-Stokes processes,
\begin{equation}
\label{ }
\bigg[\partial_t+v_s \partial_z\! - \! \frac{\chi \Gamma_0 v_s}{4 L}  G \bigg]\bar{A}_s(z,t)\!= \!- i g_0^* \bar{A}_p \bar{B}^\dag_{th}(z,t).
\end{equation}
The solution for the Stokes field is given by 
\begin{align}
\label{ }
\bar{A}_s(z,t) = - i g_0^*  \int_0^t d\tau e^{\frac{\chi \Gamma_0 v_s}{4 L}  G(t-\tau)} & \bar{A}_p \bar{B}_{th}^\dag(z-v_s(t-\tau),\tau)\nonumber \\ 
&\times \theta(z-v_s(t-\tau))
\end{align}
where $\bar{A}_s(0,t) = 0$. 
By using the correlation properties of the thermal phonon envelope
\begin{equation}
\label{ }
\langle \bar{B}_{th}(z,t) \bar{B}_{th}^\dag(z',t') \rangle = (\bar{n}_{th}+1) 
e^{-i\Delta(t-t')-\frac{\Gamma_0}{2}|t-t'|}\delta(z-z') 
\end{equation}
the simplified form of the correlation function below can be obtained
\begin{align}
\label{As-Correlation}
& \langle \bar{A}^\dag_s(z+z',t+t') \bar{A}_s(z',t') \rangle = 
- \frac{2|g_0|^2 |\bar{A}_p|^2 L}{ \Gamma_0 v_s^2 {\rm Re} (\chi) G} (\bar{n}_{th}+1)
\nonumber 
\\
& \times e^{i\Delta\left( t-\frac{z}{v_s} \right)-
\frac{\Gamma_0}{2}\left| t-\frac{z}{v_s} \right|} \nonumber \\
&
\times e^{\frac{-i{\rm Im}(\chi)\Gamma_0 G}{4 L} z}  
  \bigg[ 
e^{\frac{{\rm Re} (\chi)\Gamma_0 G}{4 L}|z|}
-
e^{\frac{{\rm Re} (\chi)\Gamma_0 G}{4 L}(z+2z')}
\bigg].  
\end{align}
This expression describes the temporal and spatial correlations in spontaneously scattered Stokes light. In particular, when the single pass gain is negative, Eq. \eqref{As-Correlation} shows that the spatial and temporal correlation length of spontaneously scattered Stokes light is limited by the phonon lifetime. This behavior shows that noise initiated stimulated emission cannot occur when $G<0$; in contrast when the single pass gain is positive, the system's gain can overcome the system's losses so that the coherence length of the emitted Stokes light becomes large. 
In addition, this expression can be used to compute the power spectrum in a variety of experimental scenarios. 
\begin{figure}[h!]
\centerline{
\includegraphics[width=9.0cm]{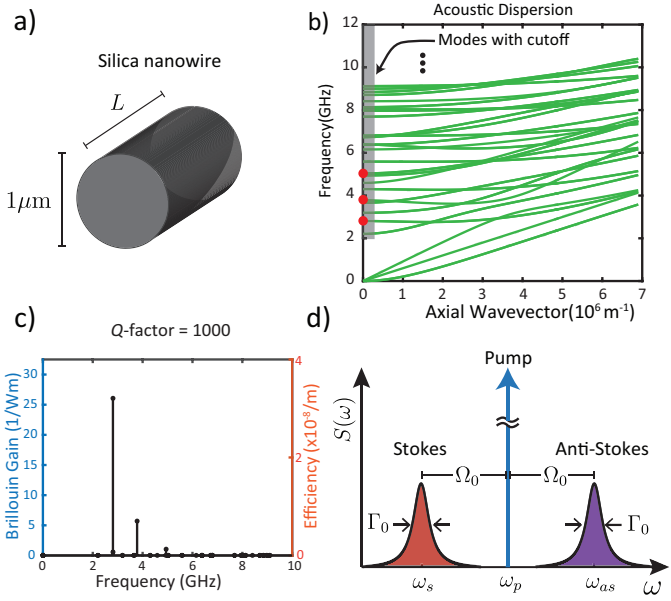}}
\caption{Spontaneous forward scattering efficiency calculation for a tapered fiber geometry. (a) A tapered fiber of diameter 1 $\mu m$ that is routinely used in quantum optics measurements. b) Acoustic dispersion curves generated numerically using finite element simulation to predict frequency range of acoustic modes that are responsible for spontaneous forward Brillouin scattering in this waveguide. c) Forward Brillouin gain coefficient, $G_B$, for acoustic modes calculated numerically using the overlap integrals and assuming a constant quality factor of 1000 for the acoustic modes. The same plot shows total spontaneous forward scattering per unit length for the Stokes light using the equation (\ref{stimtospont}). d) Schematic representation of the power spectrum of the noise, which is a lorenztian with full width at half maximum of $\Gamma_0.$ The area under the noise spectrum is integrated to get the total spontaneously scattered light.}
\label{fig2}
\end{figure}

For spontaneous noise measurements the power spectrum of the noise at position $L$ along the waveguide is defined as $S_s(\omega) = (\hbar\omega_s v)\int^{\infty}_{-\infty} d t^{\prime} e^{-i\omega t^{\prime}} \left\langle {A}_s(L,t+t^{\prime}) {A}_s^{\dagger}(L,t) \right\rangle$, where $\omega$ is measured relative to $\omega_s$ \cite{raymer1981stimulated}.  For $t \rightarrow \infty$, assuming $G \approx 0$, $S_s(\omega)$, using equation (\ref{As-Correlation}), evaluates to:
\begin{align}
S_s(\omega) \simeq  \frac{}{} \frac{4|g_0|^2 P_p (\bar{n}_{\text{th}}+1)L}{v^2\Gamma_0}\frac{(\Gamma_0/2)^2}{\omega^2+(\Gamma_0/2)^2}
\end{align} 

As an example calculation of spontaneous forward Brillouin noise we look at a tapered optical fiber that is used in quantum optics experiment (see Fig. \ref{fig2}(a)) at room temperature. Calculation of acoustic dispersion curves for this cylindrical geometry with 1 $\mu m$ diameter using numerical methods gives us the range of frequencies for slow-group velocity modes in this system (i.e. greater than 2 GHz) (see Fig. \ref{fig2}(b)). However, only the acoustic modes with large acousto-optic coupling scatter pump light to forward propagating stokes and anti-Stokes (see Fig. \ref{fig2}(c)). For instance, an acoustic mode with $\Omega_0 = 2 \pi \times 2.81 $ GHz, a Brillouin gain coefficient of $G_\text{B} = 25.9$ $ \text{W}^{-1}\text{m}^{-1}$, a $Q-$factor of 1000, and interacts with pump light at $\omega_p = 2\pi \times 194$ THz, resulting in the forward stokes scattering efficiency is $E_F/L = 3.2 \times 10^{-8} \text{m}^{-1}.$ Therefore, the \textit{total} spontaneously scattered Stokes power in a narrow band around $\omega_s$ (see Fig. \ref{fig2}(d)) in a meter long tapered fiber with 100 mW pump power is $P_s = 1/(2\pi) \int{\text{d}\omega \ S_s(\omega)} = P_p\times E_F = 3.2 \ \text{nW}.$

\begin{figure}[h!]
\centerline{
\includegraphics[width=9.0cm]{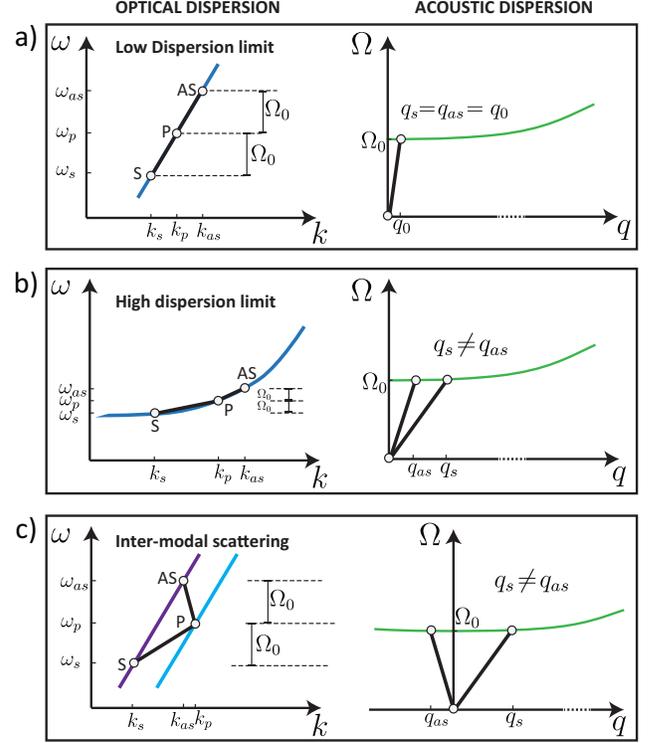}}
\caption{A schematic comparing different limits of forward Brillouin scattering. a) In the case of low optical dispersion, the optical group velocity is approximately constant, meaning both the stokes and anti-Stokes wavectors are equal (within wavevector certainty because of the finite length of the waveguide) (i.e. $q_a = q_{as} = q_o$). b) For waveguides with strong optical dispersion, group velocity is no longer the same for Stokes and anti-Stokes field. However, phonon frequency for the Stokes and anti-Stokes process are equal to $\Omega_0$ to an excellent approximation because the acoustic dispersion curve is nearly flat. Therefore, acoustic wavevector for Stokes scattering is not equal to that for anti-Stokes scattering (i.e. $q_s \neq q_{as}$). c) For inter-modal scattering, light is scattered between modes with different dispersion curves. Therefore, even for the case when those two curves have low dispersion, the phonon wavevector for Stokes and anti-Stokes are not equal (i.e. $q_s \neq q_{as}$).} 
\label{fig3}
\end{figure}

\subsection{\label{sec:level7} Symmetry breaking in forward Brillouin scattering}
In the previous section, we discussed how the noise properties for forward Brillouin scattering are different than that for backward Brillouin scattering. Contrasting behavior occurs in forward Brillouin scattering because the optical dispersion is weak and the same phonon mode couples to both the Stokes and the anti-Stokes fields (see Fig.\ref{fig3}(a)). However, there are forward Brillouin systems where this degeneracy is broken. This is possible with high optical dispersion or inter-modal scattering. {\bf cite []} In such systems, spontaneous forward Brillouin is similar to backward Brillouin scattering. We discuss this in the next two sections.

\subsubsection{\label{subsec:level8} High optical dispersion limit}
In highly dispersive waveguides, such as slow-light photonic crystal and Bragg waveguides \cite{qiu2012stimulated, krauss2007slow, povinelli2004slow, joannopoulos2011photonic, oskooi2009zero}, enhanced group velocity dispersion for the optical modes (see Fig. \ref{fig3}(b)) can produce an appreciable difference in the Stokes and Anti-Stokes acoustic wavevector. This can be seen from $q_s \approx \Omega_0/v_g(\omega_p)$ and $q_{as} \approx \Omega_0/v_g(\omega_{as}).$ Such a difference in wavevectors can only be resolved in sufficiently long waveguides. If $|q_{as} -q_{s}|L > \pi$, then the Stokes and anti-Stokes fields  do not couple to the same phonon mode.
 
When these conditions are met, Stokes and anti-Stokes processes can be treated separately.  Following a similar approach to that outlined in the previous section and ignoring the dynamics of the anti-Stokes field we get the following spontaneously scattered stokes signal in the long time limit $\tau \rightarrow \infty$ at position $z=L$
\begin{align} \label{EquationwithsymmetryBreaking}
\resizebox{1.0\hsize}{!}{$\left\langle |\bar{A}_s(L)|^2 \right\rangle \!=\! \frac{|g_0|^2 |\bar{A}_p|^2 L}{v^2} (\bar{n}_{\text{th}}\!+\!1)e^{\frac{G}{2}} \left( I_0\left(G/2\right)\!-\!I_1\left(G/2\right) \right)$},
\end{align}
where  
\begin{align}
G = \frac{4|g_0|^2L |\bar{A}_p|^2}{v\Gamma_0} = G_B P_p L.
\end{align}
In contrast to the results of the previous section, $G$ is positive, indicating that noise is exponentially amplified in highly dispersive systems. The dynamics and resulting noise properties in this system are similar to that in backward Brillouin scattering. This is due to the symmetry breaking between the Stokes and the anti-Stokes processes and the assumption that phonons are local in forward Brillouin scattering (i.e. $\partial \bar{B}/\partial z \approx 0$). Finally, since the single-pass gain $G=G_B P_P L$ can be large, we expect the exponential growth of spontaneously scattered light to initiate stimulated forward Brillouin scattering in this system.

\subsubsection{\label{subsec:level9} Inter-modal scattering}
In contrast to intra-modal scattering, inter-modal scattering involves scattering of light between two distinct optical bands (see Fig. \ref{fig3}(c)). From the figure it is clear that even for modes with little optical dispersion, the Stokes and the anti-Stokes acoustic wavevector can be different in inter-modal scattering. In fact, the phonon wavevectors can be counter propagating. Again, this phase-matching requirement breaks the degeneracy between Stokes and anti-Stokes processes. The resulting dynamics and noise properties are the same as for the high-optical dispersion limit discussed above. 

\section{\label{sec:discconc}Discussion and Conclusion}
In summary, we presented a generalizable quantum mechanical theory of forward Brillouin scattering. This approach incorporates the spatial dynamics and highlights the traveling nature of the waves involved in Brillouin interactions. The Heisenberg equations of motion were used to calculate the stimulated amplification of Stokes light through forward Brillouin scattering. Spontaneous scattering, resulting from thermal fluctuations of guided acoustic phonons, was calculated by adding dissipation and a Langevin driving force to the equation of motion for the phonon field. 

The coupling strength, which takes into account both electrostriction and radiation pressure, can be calculated for arbitrary waveguide geometry. This allowed us to derive analytical expressions for forward scattering efficiency for any waveguide, which could be useful in predicting and understanding noise in many quantum optics experiments. In addition, we showed that spontaneously scattered Stokes can be calculated knowing the Brillouin gain coefficient obtained from stimulated measurements, unifying the treatment of spontaneous (formerly studied as GAWBS) with the stimulated forward Brillouin scattering.

We also showed that for intra-modal scattering in the non-dispersive waveguide spontaneously scattered light grows linearly with device length. This behavior is markedly different than that for backward Brillouin scattering where noise grows exponentially, allowing noise to initiate stimulated Brillouin scattering. This difference arises from the fact that, in forward Brillouin scattering, phonons can simultaneously phase match to both the Stokes and the anti-Stokes fields. However, this degeneracy  is broken for the case of highly dispersive systems or inter-modal scattering, leading to noise properties similar to backward Brillouin scattering.

Interestingly, we demonstrated that in the undepleted pump regime the stimulated Stokes signal grows quadratically with both length and pump power whereas the spontaneously scattered Stokes signal (i.e. spontaneous noise) grows linearly with length and pump power for intra-modal scattering. These distinct behaviors suggest that forward Brillouin amplification may have surprising benefits as further signal processing applications are developed based on such interactions. 

\section{\label{sec:acknow} Acknowledgments}
Primary support for this work was provided by NSF MRSEC DMR-1119826. This work was also supported by the MesoDynamic Architectures  program  at  DARPA under the direction of Dr. Daniel Green and Yale University. The authors thank Florian Marquardt, Heedeuk Shin, and Eric Kittlaus for useful technical discussions involving optomechanical processes and Brillouin interactions.

\appendix 
\section{}
In this appendix we follow the approach outlined by Sipe \emph{et al.}  \cite{sipe2015hamiltonian} and express the acousto-optic Hamiltonian in terms of envelope operators. Let us consider a waveguide segment of length $L$ that is axially invariant in $z$ direction and supports both acoustic and optical modes. The complete opto-acoustic Hamiltonian that takes into account all possible interactions between light and sound  for this system is given by \cite{sipe2015hamiltonian}
\begin{multline}
H = \int{\frac{\pi^i(\textbf{r})\pi^i(\textbf{r})}{2\rho(\textbf{r})}} \text{d}\textbf{r} + \frac{1}{2}\int{S^{ij}(\textbf{r})c^{ijkl}(\textbf{r})S^{kl}(\textbf{r}) \text{d}\textbf{r}}+\\
\frac{1}{2\mu_o}\int{B^i(\textbf{r})B^i(\textbf{r}) \text{d}\textbf{r}}+\frac{1}{2\epsilon_o}\int{D^i(\textbf{r})\beta^{ij}(\textbf{r})D^i(\textbf{r})\text{d}\textbf{r}}.
\end{multline}
Here, $\pi$\textbf{(r)} is the conjugate momenta of the acoustic displacement field operator \textbf{u(r)}, $\rho(\textbf{r})$ is the density, $c^{ijkl}(\textbf{r})$ is the elastic constant tensor, $S^{ij}(\textbf{r})=1/2(\partial u^i\textbf{(r)}/\partial r^j+\partial u^j\textbf{(r)}/\partial r^i)$ is the strain operator, $\textbf{D(r)}$ is the electric displacement field operator, $\textbf{B(r)}$ is the magnetic field operator and $\epsilon_r^{ij}\textbf{(r)}=1/\beta^{ij}\textbf{(r)}$ is the relative dielectric constant tensor. 

For a long waveguide segment (i.e. $L \rightarrow \infty$), the acoustic displacement operator \textbf{u(r)} and the electric displacement operator \textbf{D(r)} can be written using the normal mode expansion as follows
\begin{align}
\textbf{u(r)} &= \sum_{\alpha} \int{\frac{\text{d}q}{\sqrt{2\pi}} \sqrt{\frac{\hbar \Omega_{\alpha q}}{2}}b_{\alpha q} \boldsymbol{\mathscr{U}}_{\alpha q}(\textbf{r}_\perp)e^{iqz} + \text{H.c.}} \\
\textbf{D(r)} &= \sum_{\gamma} \int{\frac{\text{d}k}{\sqrt{2\pi}} \sqrt{\frac{\hbar \omega_{\gamma k}}{2}} a_{\gamma k} \textbf{d}_{\gamma k}(r_{\perp})e^{ikz}+ \text{H.c.}}
\end{align}
Here, $b_{\alpha q}$ and $a_{\gamma k}$ above represent the acoustic mode amplitude operator and the optical amplitude operator for a mode with transverse profile and longitudinal wavenumber given by $(\boldsymbol{\mathscr{U}}_{\alpha q} (\textbf{r}_\perp), q)$ and $(\textbf{d}_{\gamma k}(\textbf{r}_\perp), k)$ respectively. $\Omega_{\alpha q}$ and $\omega_{\gamma k}$ are the acoustic and optical frequencies respectively. 
The transverse modes are normalized such that
\begin{align}
\Omega^2_{\alpha q} \int{\text{d}\textbf{r}_{\perp} \rho(\textbf{r}_{\perp})\boldsymbol{\mathscr{U}}^*_{\alpha q}(\textbf{r}_\perp)\cdot \boldsymbol{\mathscr{U}}_{\alpha q}(\textbf{r}_\perp)} &= 1, \\
\frac{1}{\epsilon_o} \int{\text{d}\textbf{r}_{\perp} \beta_r(\textbf{r}_{\perp}) \textbf{d}^*_{\gamma k}(\textbf{r}_\perp) \cdot \textbf{d}_{\gamma k}(\textbf{r}_\perp)} &= 1,
\end{align}
and the mode operators satisfy the following commutation relations:
\begin{align}
[b_{\alpha q},b_{\alpha^{\prime} q^{\prime}}] &= 0; \ [b_{\alpha q},b^{\dagger}_{\alpha^{\prime} q^{'}}] = \delta_{\alpha \alpha^{\prime}} \delta(q-q^{\prime}) \\
[a_{\gamma k},a_{\gamma^{\prime} k^{\prime}}] &= 0; \ [a_{\gamma k},a^{\dagger}_{\gamma^{\prime} k^{\prime}}] = \delta_{\gamma \gamma^{\prime}} \delta(k-k^{\prime}). 
\end{align}
The quantized version of this  Hamiltonian in terms of the mode amplitude operators, neglecting the vacuum fluctuations and writing the dominant interaction terms between the photons and the phonons, is 
\begin{align} \label{Hfull}
&H = H^{A} + H^{EM} + V
\end{align}
\begin{widetext}
 $H= \sum_{\alpha} \int {\text{d}q \ \hbar \Omega_{\alpha q} b_{\alpha q}^{\dagger}b_{\alpha q}}+\sum_{\gamma} \int {\text{d}k \ \hbar \omega_{\gamma k} a_{\gamma k}^{\dagger}a_{\gamma k}}+\sum_{\alpha,\gamma,\gamma^{\prime}} \int{\frac{\text{d}k \text{d}k^{\prime}\text{d} q}{(2\pi)^{\frac{3}{2}}} \left( \ a_{\gamma k}^{\dagger} a_{\gamma^{\prime}k^{\prime}}b_{\alpha q} \int{\text{d}z \ g(\gamma k;\gamma^{\prime}k^{\prime};\alpha q) e^{i(k^{\prime}-k+q)z}}+\text{H.c.}\right)}.$
\end{widetext}
The coupling term, $g(\gamma k;\gamma^{\prime}k^{\prime};\alpha q)$, for the process involving annihilation of a photon to give a photon and a phonon is given by 
\begin{widetext}
$g(\gamma k;\gamma^{\prime}k^{\prime};\alpha q) = \frac{1}{\epsilon_o}\sqrt{\frac{\hbar \omega_{\gamma k}}{2}} \sqrt{\frac{\hbar \omega_{\gamma^{\prime} k^{\prime}}}{2}} \sqrt{\frac{\hbar \Omega_{\alpha q}}{2}} \int{\text{d}\textbf{r}_{\perp} (d_{\gamma k}^i(\textbf{r}_\perp))^*d_{\gamma^{\prime} k^{\prime}}^j(\textbf{r}_\perp) \left( p^{ijlm}(\textbf{r}_\perp) \frac{\partial \mathscr{U}_{\alpha q}^l (\textbf{r}_\perp)}{\partial r^m}-\delta^{ij} \left(\frac{\partial \beta_{r}(\textbf{r}_\perp)}{\partial r^l} \right)\mathscr{U}_{\alpha q}^l (\textbf{r}_\perp)\right)}.$
\end{widetext} 
The first term in the right, which is the transverse mode profile overlap between the optical and acoustic modes given by $\textbf{d}_{\gamma k}(\textbf{r}_\perp)$ and $\boldsymbol{\mathscr{U}}_{\alpha q} (\textbf{r}_\perp)$ respectively, represents the coupling strength of the photo-elastic interaction whereas the second term, which is the displacement induced change in relative dielectric index, represents the coupling strength due to radiation pressure as well as moving boundaries; $p^{ijlm}$ is the photo-elastic tensor of the medium.
At this point we introduce the envelope field operators to represent acoustic(optical) excitation with a given spatial mode $\alpha (\gamma)$ that is centered around some wavenumber $q_j(k_j)$:
\begin{align} \label{phonEnvelop1}
B_{\alpha j}(z,t) &= \int{ \frac{\text{d}q}{\sqrt{2\pi}} b_{\alpha q}(t) e^{i(q-q_j)z}} \\ \label{phononEnvelop2}
A_{\gamma j}(z,t) &= \int{\frac{\text{d}k}{\sqrt{2\pi}} a_{\gamma k}(t) e^{i(k-k_j)z}}
\end{align}
The equal time commutation relation for envelope field operators can be derived from the commutation relations for the mode operators and are given as follows:
\begin{align} \label{commuModeEnvelop1}
[B_{\alpha j}(z,t),B^{\dagger}_{\alpha^{\prime} j^{\prime}}(z^{\prime},t)] &= \delta_{\alpha \alpha^{\prime}} \delta_{j j^{\prime}} \delta(z-z^{\prime}), \\ \label{commuModeEnvelop2}
[A_{\gamma j}(z,t),A^{\dagger}_{\gamma^{\prime} j^{\prime}}(z^{\prime},t)] &= \delta_{\gamma \gamma^{\prime}} \delta_{j j^{\prime}} \delta(z-z^{\prime}).   
\end{align}
Assuming the optical or acoustic excitations are narrow-band so that the excitation frequencies and the transverse mode profiles remain constant around the carrier wavenumbers the elastic displacement and the electric displacement can be expressed in terms of the envelope operators as

\begin{align}
\textbf{u(r},t)&\simeq \sum_{\alpha j}\left( \sqrt{\frac{\hbar \Omega_{\alpha j}}{2}} \boldsymbol{\mathscr{U}}_{\alpha q_j}(\textbf{r}_\perp) B_{\alpha j}(z,t) e^{iq_jz} +\text{H.c.} \right)\\
\textbf{D(r},t)&\simeq \sum_{\gamma j}\left( \sqrt{\frac{\hbar \omega_{\gamma j}}{2}} \textbf{d}_{\gamma k_j}(\textbf{r}_\perp) A_{\gamma j}(z,t) e^{ik_jz} +\text{H.c.} \right).
\end{align}
Here, the sum over $j$ represents the sum over all the acoustic and optical excitation in the waveguide segment.

To write the Hamiltonian in terms of envelope field operators we first Taylor expand the phonon frequency $\Omega_{\alpha q}$ and the photon frequency $\omega_{\gamma k}$ around the carrier $j$:
\begin{align} \label{Omegaexp}
\Omega_{\alpha q} &= \Omega_{\alpha j}+ (q-q_j)\frac{\partial \Omega_{\alpha q}}{\partial q}\biggr\rvert_{q=q_j}+ \dots \\\label{omegaexp}
\omega_{\gamma k} &= \omega_{\gamma j}+ (k-k_j)\frac{\partial \omega_{\gamma k}}{\partial k}\biggr\rvert_{k=k_j}+ \dots.
\end{align}
, where $\text{v}_{\alpha j} = \partial\Omega_{\alpha q}/\partial q \rvert_{q=q_j}$ is the acoustic group velocity and $v_{\gamma j} = \partial\omega_{\gamma k}/\partial k \rvert_{k=k_j}$ is the optical group velocity.

Substituting (\ref{Omegaexp}) and (\ref{omegaexp}) into the expression for $H$ in  (\ref{Hfull}) and using the relations (\ref{phonEnvelop1}) and (\ref{phononEnvelop2}) we can write the Hamiltonian in terms of the envelope field operators
\begin{align*}
H^{A} & = \sum_{\alpha j}\biggr (\hbar \Omega_{\alpha j} \int{\text{d}z \ B^{\dagger}_{\alpha j}(z,t)B_{\alpha j}(z,t)} -\\
& i\hbar \text{v}_{\alpha j} \int{\text{d}z \ B^{\dagger}_{\alpha j}(z,t) \frac{\partial B_{\alpha j}}{\partial z}(z,t)} +\dots \biggr ), \\
H^{EM} & = \sum_{\gamma j} \biggr( \hbar \omega_{\gamma j} \int{\text{d}z \ A^{\dagger}_{\gamma j}(z,t)B_{\gamma j}(z,t)} -\\
& i\hbar v_{\gamma j} \int{\text{d}z \ A^{\dagger}_{\gamma j}(z,t) \frac{\partial A_{\gamma j}}{\partial z}(z,t)} +\dots \biggr)\\
V &= \sum_{\alpha,\gamma, \gamma^{\prime}}\sum_{j,j^{\prime},l} \biggr ( g(\gamma k_j;\gamma^{\prime}k_{j^{\prime}};\alpha q_l) \delta(k_{j^{\prime}}-k_j+q_l) \times \\
&\int{\text{d}z \ A^{\dagger}_{\gamma j}(z,t) A^{}_{\gamma^{\prime} j^{\prime}}(z,t)B_{\alpha l}(z,t)} +\text{H.c.} \biggr ).
\end{align*}
In deriving the interaction term in the Hamiltonian, $V$, we have taken the coupling strength $g(\gamma k_j;\gamma^{\prime}k_{j^{\prime}};\alpha q_l)$ out of the spatial integral assuming that the coupling strength is constant over narrow bands around carrier wavenumbers.

Finally, the time evolution of the envelope fields are then given by Heisenberg equation of motion 
\begin{align}
\frac{\partial B_{\alpha j}(z,t)}{\partial t} &= \frac{1}{i\hbar}[B_{\alpha j}(z,t),H] \\
\frac{\partial A_{\gamma j}(z,t)}{\partial t} &= \frac{1}{i\hbar}[A_{\gamma j}(z,t),H].
\end{align}
The equal time commutator relations in (\ref{commuModeEnvelop1}) and (\ref{commuModeEnvelop2}) can then be used to find the coupled mode equations for the envelope fields. If we ignore the terms corresponding to the higher order dispersion in the Hamiltonian, then it amounts to making slowly varying envelope approximation for the envelope fields.
\appendixpageoff

\bibliographystyle{apsrev4-1} 
\bibliography{mybib3}
\end{document}